\documentclass[conference]{IEEEtran}
\IEEEoverridecommandlockouts

\usepackage{cite}
\usepackage{amsmath,amssymb,amsfonts}
\usepackage[utf8]{inputenc}
\usepackage{algorithmic}
\usepackage{graphicx}
\usepackage{textcomp}
\usepackage{xcolor}
\usepackage{verbatim}   
\usepackage{url}
\def\BibTeX{{\rm B\kern-.05em{\sc i\kern-.025em b}\kern-.08em
    T\kern-.1667em\lower.7ex\hbox{E}\kern-.125emX}}
\usepackage[linesnumbered, ruled, boxed]{algorithm2e}
\usepackage{float}
\usepackage{enumitem} 

\begin{document}

\title{Memory-Augmented Log Analysis with \\Phi-4-mini: Enhancing Threat Detection in Structured Security Logs}

\author{
\IEEEauthorblockN{Anbi Guo, Mahfuza Farooque}
\IEEEauthorblockA{School of Electrical Engineering and Computer Science,\\
Pennsylvania State University, University Park, PA, USA \\
\{aqg6077, mff5187\}@psu.edu}
}

\maketitle

\begin{abstract}
Structured security logs are critical for detecting advanced persistent threats (APTs). Large language models (LLMs) struggle in this domain due to limited context and domain mismatch. We propose \textbf{DM-RAG}, a dual-memory retrieval-augmented generation framework for structured log analysis. It integrates a short-term memory buffer for recent summaries and a long-term FAISS-indexed memory for historical patterns. An instruction-tuned Phi-4-mini processes the combined context and outputs structured predictions. Bayesian fusion promotes reliable persistence into memory. On the UNSW-NB15 dataset, DM-RAG achieves 53.64\% accuracy and 98.70\% recall, surpassing fine-tuned and RAG baselines in recall. The architecture is lightweight, interpretable, and scalable, enabling real-time threat monitoring without extra corpora or heavy tuning.
\end{abstract}


\section{Introduction}
Structured security logs, such as those from network traffic monitors, intrusion detection systems, or system audits, form the basis for detecting and responding to cyber threats. Despite the success of large language models (LLMs) in natural language tasks, they often struggle to generalize to specific domains such as Biomedicine~\cite{gu2021domain,luo2022biogpt}, Finance~\cite{wu2023bloomberggpt}, Medicine~\cite{singhal2025toward}, and Security~\cite{google2024rsa} without adaptation, due to differences in syntax, semantics, and behavioral patterns.

This stems from the lack of interpretability in LLM outputs~\cite{gilpin2018explaining} and the gap between their pre-training corpora and security logs~\cite{arazzi2025nlp}. Complex persistent threats, such as advanced persistent threats (APTs), often manifest as multistage attacks. Their key log events are dispersed across long periods~\cite{hutchins2011killchain,chen2014apt,mitre2024attack}, challenging log analysis methods based on fixed context windows. LLMs also face token context limits—information beyond the window may be discarded or compressed. Reasoning over log entries across multiple windows and days thus becomes difficult~\cite{liu2023lost}.

To address this, we introduce a ``memory'' mechanism into log analysis, based on instruction tuning and adaptive real-time RAG technology.

This paper uses Phi-4-mini~\cite{abouelenin2025phi4mini}, a lightweight, open-source, decoder-only LLM as the foundation. We simulate working memory and long-term memory from human cognition~\cite{cowan2008memory}, used respectively to retain recent behavior and persistent attack patterns. Two adaptive memories are automatically maintained to assist in analysis and reasoning on structured logs. To enhance cross-temporal reasoning, instruction tuning is applied to train the model for memory usage. A rolling summarization mechanism captures and maintains recent behaviors. In addition, a long-term retrievable memory is built with RAG to support integration of historical patterns. This architecture strengthens the model's ability to combine local context with global historical signals.

Specifically, DM-RAG maintains two complementary memory streams. The first is a rolling summary to retain recent behaviors, while the second is a real-time updated RAG that stores high-confidence behaviors.

We evaluated our system on the UNSW-NB15 dataset~\cite{moustafa2015unsw}, split into training and test sets. We compared it with three configurations: the original Phi-4-mini, a Phi-4-mini fine-tuned with Low-Rank Adaptation (LoRA)~\cite{Edward2022Lora}, and a RAG-enhanced Phi-4-mini incorporating external threat knowledge. Experimental results show that DM-RAG achieves the highest recall (98.70\%) and F1 score (69.59\%), validating its effectiveness for high-coverage, multistage threat detection in structured logs.

\section{Related Work}

\subsection{Language Modeling }

Deep learning has become increasingly popular for log anomaly detection recently. One of the earliest models is \textit{DeepLog}~\cite{du2017deeplog}, which treats system logs as natural language sequences and uses a Long Short-Term Memory (LSTM) network to model log key sequences for anomaly detection.

With advances in natural language processing (NLP), researchers explored transformer-based models for log analysis. \textit{LogBERT}~\cite{guo2021logbert} was the first to use a bidirectional transformer encoder to learn contextual relationships between log keys. It performs \textit{Masked Log Key Prediction} to identify expected log entries and applies \textit{Volume of Hypersphere Minimization} to analyze log representations, where normal logs cluster near the hypersphere center while anomalies deviate.

Building on this, \textit{LogGPT}~\cite{han2023loggpt} introduced a reinforcement learning (RL)-augmented transformer with a reward mechanism for correct predictions. \textit{LogLLaMA}~\cite{yang2025logllama} further refined this approach by rewarding predictions based on the Top-K most probable candidates.

Beyond architectural innovations, \textit{SecEncoder}~\cite{bulut2024secencoder} takes a different approach. It forgoes pre-trained models and trains entirely from scratch on raw logs, focusing on log-specific semantics without pretraining cost.

\subsection{Retrieval-Augmented Generation (RAG)}
RAG combines language models with external retrieval to expand context and grounding. While effective in knowledge-intensive tasks (e.g., QA, summarization), it assumes a static knowledge base and cannot adapt to evolving log events without frequent re-indexing. Performance of RAG-seq improves as K increases, with the best effect for tokens when K=10~\cite{lewis2020rag}. More recently, \textit{RagLog}~\cite{pan2024raglog} applied RAG for log anomaly detection through a QA pipeline, leveraging external log contexts to generate and evaluate expected log entries. This improves interpretability and gives the model broader contextual knowledge.

\subsection{Memory-Augmented Language Models}
To overcome fixed context limits, recent work augments LLMs with external memory. MemGPT~\cite{packer2023memgpt} introduces an OS-inspired hierarchical memory architecture, enabling LLMs to manage virtual context via function calls. It separates in-context memory from out-of-context storage, and lets the model retrieve, store, and update relevant information autonomously. This supports long-term reasoning in tasks such as multi-session dialogue and document analysis, improving over fixed-context baselines.

\subsection{Core Ideas in Log Anomaly Detection}
Despite diverse approaches, the core idea is consistent: modeling contextual dependencies of log sequences with transformers and detecting anomalies based on prediction confidence. If an observed log entry is not in the model's Top-K predictions given its context, it is considered anomalous due to its low probability under the learned log model.

\section{Method}

\subsection{Overview}
We introduce a Dual-Memory Retrieval-Augmented Generation (DM-RAG) system for sequential log analysis. Our system wraps a compact decoder-only language model (Phi-4-mini) with two interacting memory modules supporting continual reasoning and anomaly classification. Each inference window processes log entries to generate a summary and prediction.

Two memory structures are maintained:

\textbf{Short-Term Memory (STM):} A rolling buffer of size $K=10$ that retains recent summaries with confidence scores.

\textbf{Long-Term Memory (LTM):} A persistent vector-based store indexed via semantic embeddings, storing high-confidence summaries for retrieval and reuse.

These memories are injected into subsequent prompts to enable retrieval-augmented reasoning with both recent and accumulated context.

\begin{figure}
    \centering
    \includegraphics[width=0.48\textwidth]{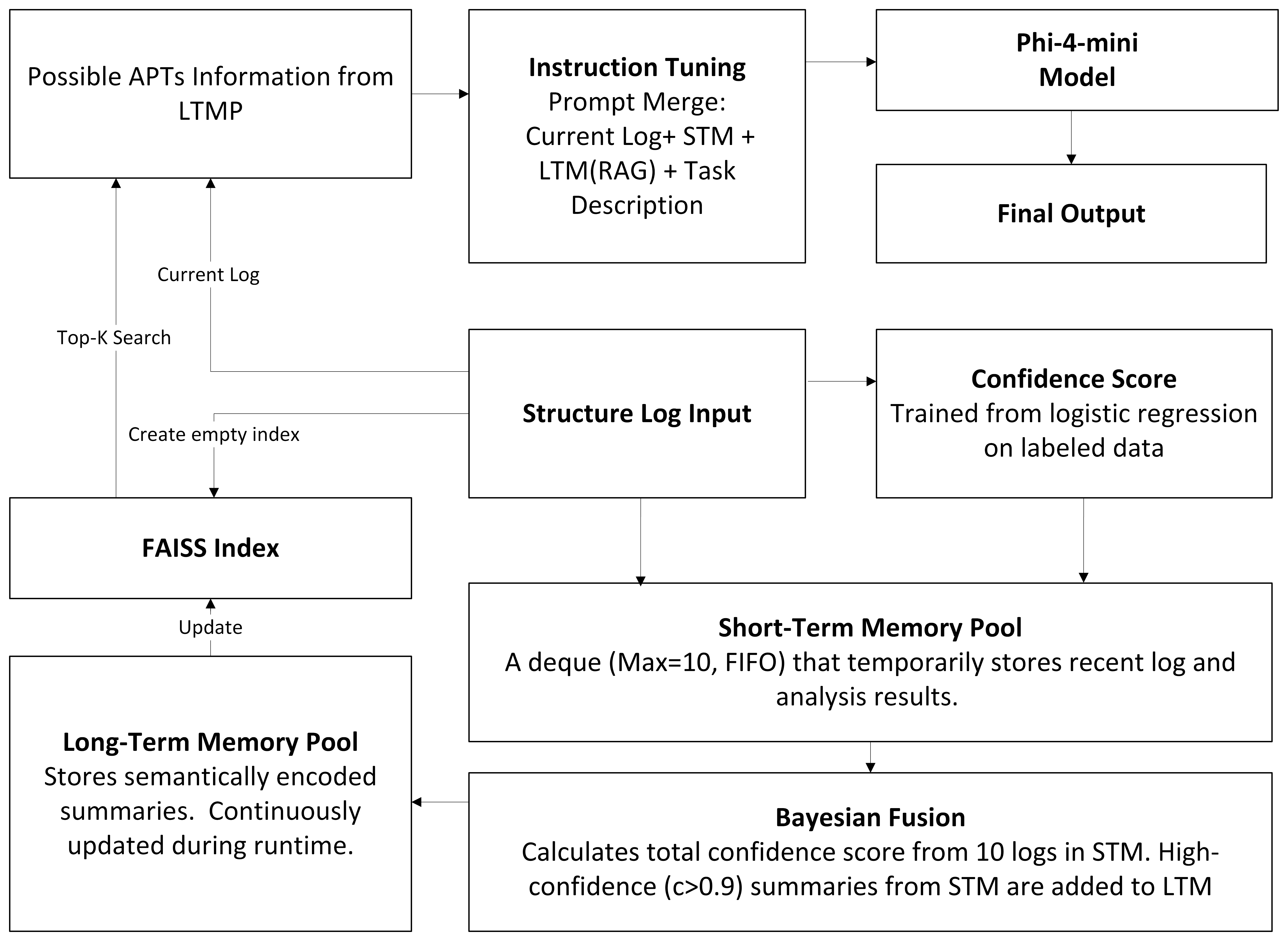}
    \caption{Incoming network logs are analyzed by an instruction-tuned LLM with dual memory. The short-term memory stores recent summaries and scores, while the long-term memory retrieves relevant high-confidence examples via FAISS. These jointly inform the prompt to Phi-4-mini for threat reasoning and classification. STM's confidence score are periodically compressed with Bayesian fusion, and high-confidence results are promoted to LTM for future retrieval.}
    \label{fig:your_label}
\end{figure}

\subsection{Initial Confidence Generation via Logistic Regression}

Logistic regression is widely used in anomaly detection. It estimates the probability of normal or anomalous from the input vector~\cite{wu2023effectiveness}.

To estimate an initial confidence score for each log entry, we train a logistic regression model on a large labeled dataset. The process is as follows.

First, features are extracted, including flow-level metrics like byte and packet counts, flow duration, traffic rate, TTL, TCP indicators, jitter, and application-layer statistics. All continuous features are normalized to $[0, 1]$ using min-max scaling for consistency across dimensions:
\begin{equation}
x_i^{\text{norm}} = 
\frac{x_i - \min(x_i)}{\max(x_i) - \min(x_i)}
\label{eq:normalization}
\end{equation}

Next, the labeled UNSW-NB15 dataset is prepared. Each instance consists of a normalized feature vector $\mathbf{x}^{(i)}$ and a binary label $y^{(i)} \in \{0, 1\}$ indicating normal or anomalous:
\begin{equation}
D = \left\{ \left( \mathbf{x}^{(i)}, y^{(i)} \right) \right\}_{i=1}^{N}
\label{eq:dataset}
\end{equation}

We then train the model to map feature vectors to anomaly probabilities. The posterior probability of anomaly is:
\begin{equation}
P(y = 1 \mid \mathbf{x}) = 
\frac{1}{1 + \exp\!\left(-\mathbf{w}^\top \mathbf{x} - b\right)}
\label{eq:logistic}
\end{equation}

For a new input $\mathbf{x}$, the model outputs a probability $p \in [0, 1]$, interpreted as the initial anomaly confidence score:
\begin{equation}
\text{score}(\mathbf{x}) = P(y = 1 \mid \mathbf{x})
\label{eq:score}
\end{equation}

\subsection{Memory Generation}

When a new log entry is received, the system performs memory generation.

First, the log is encoded into a vector using SentenceTransformer MiniLM-L6-v2~\cite{wang2020minilm} and retrieves relevant summaries from long-term memory (LTM) via FAISS~\cite{johnson2019billion}. These serve as background for reasoning.

Next, the log entry is embedded into a structured prompt and passed to a language model, which outputs a natural language summary, a confidence score from 0 to 1, and an attack label such as ``Normal,'' ``DoS,'' or ``Reconnaissance.''

The output is stored in short-term memory (STM) as an object with log, analysis, confidence, and timestamp. STM is a sliding window queue of the most recent $K=10$ entries, maintaining localized temporal context without manual management.

If the confidence score exceeds a threshold ($c_i \geq 0.9$), the summary is promoted to LTM, encoded and stored in the FAISS index with metadata.

During future analysis, both STM contents and the top-10 retrieved LTM summaries are included in the reasoning prompt, providing context. This design lets the system learn from past observations, incorporate new insights, and make more accurate decisions over time.

\subsection{Memory Compression and Promotion}

When STM reaches full capacity ($K=10$), compression is triggered. The model is prompted to merge related summaries, discard redundancies, and output a merged narrative summary.

To compute a confidence score for each log entry, we adopt a Bayesian fusion framework over its structured features~\cite{yu2016bayesian,Dai2023}. Each feature indicates anomalous behavior. We model the normalized values of selected features as independent samples drawn from Beta distributions conditioned on the underlying class.

Let $\{x_i\}_{i=1}^n$ denote the set of $n$ continuous features extracted from a single network flow instance. Each $x_i$ is normalized to $[0,1]$ and modeled as a class-conditional Beta sample:

\begin{equation}
x_i \sim 
\begin{cases}
\mathrm{Beta}(\alpha_1, \beta_1), & \text{if } y = 1 \;\; \text{(anomalous)} \\
\mathrm{Beta}(\alpha_0, \beta_0), & \text{if } y = 0 \;\; \text{(normal)}
\end{cases}
\label{eq:beta-distribution}
\end{equation}

The likelihood of observing the feature set $\{x_i\}$ under each class is computed by assuming conditional independence:

\begin{equation}
P(\{x_i\} \mid y) = \prod_{i=1}^{n} \mathrm{Beta}\!\left(x_i; \alpha_y, \beta_y\right)
\label{eq:likelihood}
\end{equation}

Applying Bayes' theorem, we compute the posterior probability of the log being anomalous as:

\begin{equation}
\scalebox{0.9}{$
P(y = 1 \mid \{x_i\}) =
\frac{P(\{x_i\} \mid y = 1)\, P(y = 1)}
     {P(\{x_i\} \mid y = 1)\, P(y = 1) + P(\{x_i\} \mid y = 0)\, P(y = 0)}
$}
\label{eq:bayes-posterior}
\end{equation}

This posterior is the fused anomaly confidence score for the log entry. The Beta distribution parameters $(\alpha_y, \beta_y)$ are estimated from the training set using the ground-truth binary label $y \in \{0,1\}$.

Each log entry includes features such as basic flow metadata, packet and byte counts, traffic rate and load, TTL and packet loss indicators, timing and jitter, TCP characteristics, payload statistics, connection counters, application-level indicators, a boolean flag, and labels.

This posterior is the overall anomaly confidence score, enabling principled decision-making while accounting for contributions of heterogeneous flow features.

\subsection{Persistent Memory and Retrieval}

The Long-Term Memory (LTM) stores compressed summaries. Each summary is encoded into a 384-dimensional embedding using the all-MiniLM-L6-v2 model and indexed via FAISS. During inference, the top-10 most relevant entries are retrieved by cosine similarity and incorporated into the prompt. This provides continuity and helps detect patterns.

\subsection{Prompt Construction}

Each prompt is built from four components:

\textbf{Task Description:} Defines the objective, such as attack detection and categorization.

\textbf{Long-Term Memory (LTM):} Retrieved summaries from previous analyses.

\textbf{Short-Term Memory (STM):} Behavior summaries from recent logs.

\textbf{Current Log Entry:} The active log entry for analysis.

This composite prompt format enables instruction-following and supports few-shot reasoning with memory-based examples.

\subsection{Comparison to Prior Strategies}

Unlike traditional log parsers or static detectors, DM-RAG supports:

Continual learning through dynamic memory updates.

Interpretable natural language summaries of behavior.

Probabilistic uncertainty aggregation via Bayesian methods.

\subsection{Relation to Prior Work}

Our confidence fusion mechanism is inspired by the Bayesian ensemble model proposed in~\cite{yu2016bayesian}, first used for unsupervised anomaly detection. We adapt it to fuse confidence scores from multiple LLM-generated summaries and apply it to memory compression and promotion in our architecture.

\subsection{Algorithm}

The proposed method is described in Algorithm~\ref{alg:log_analysis}.

\begin{algorithm}
\caption{Log Analysis with Memory-Augmented RAG and Bayesian Fusion}
\label{alg:log_analysis}
\begin{algorithmic}[1]

\STATE \textbf{Input:} Structured logs $\{l_1, ..., l_n\}$; Encoder $\mathcal{E}$; 
Instruction-tuned LLM $\mathcal{G}$; FAISS index $\mathcal{M}_{LTM}$ (initially empty); 
Short-term memory buffer $\mathcal{M}_{STM}$ (size $k$); Promotion threshold $\tau$ ($0.9$)

\STATE \textbf{Output:} Analysis results with summaries and anomaly labels

\STATE \textbf{Step 1: Confidence Model Preparation}
\STATE Load labeled dataset $D = \{ (\mathbf{x}^{(i)}, y^{(i)}) \}_{i=1}^N$
\FOR{each feature vector $\mathbf{x}^{(i)}$ in $D$}
    \STATE Normalize features to $[0,1]$ using min-max scaling
\ENDFOR
\STATE Train logistic regression model $M$ to estimate $P(y=1 \mid \mathbf{x})$
\STATE Define scoring function $\text{score}(\mathbf{x}) \leftarrow M(\mathbf{x})$

\STATE \textbf{Step 2: Online Log Analysis}
\FOR{each log $l_t$ in $\{l_1, ..., l_n\}$}
    \STATE $v_t \leftarrow \mathcal{E}(l_t)$ \COMMENT{Encode current log}
    \STATE $S_{LTM} \leftarrow$ FAISS.Retrieve($v_t$, top-10) \COMMENT{Top-k retrieval}
    \STATE Build prompt $P_t \leftarrow$ Merge(Task, $S_{LTM}$, $\mathcal{M}_{STM}$, $l_t$)
    \STATE $output \leftarrow \mathcal{G}(P_t)$ \COMMENT{LLM inference}
    \STATE Parse $(summary_t, conf_t, label_t) \leftarrow output$
    \STATE $\mathcal{M}_{STM} \leftarrow \mathcal{M}_{STM} \cup \{(summary_t, conf_t)\}$

    \IF{$|\mathcal{M}_{STM}| = k$}
        \STATE Build compression prompt $P_{STM} \leftarrow$ Compress($\mathcal{M}_{STM}$)
        \STATE $merged \leftarrow \mathcal{G}(P_{STM})$
        \STATE $conf_{fused} \leftarrow$ BayesianFusion($\{conf_i\}$ in $\mathcal{M}_{STM}$)
        \IF{$conf_{fused} > \tau$}
            \STATE $v \leftarrow \mathcal{E}(merged)$
            \STATE $\mathcal{M}_{LTM} \leftarrow \mathcal{M}_{LTM} \cup \{v\}$ \COMMENT{Add to FAISS}
        \ENDIF
        \STATE Reset $\mathcal{M}_{STM} \leftarrow \{(merged, conf_{fused})\}$
    \ENDIF

    \STATE Save analysis result: $\{l_t, summary_t, label_t\}$
\ENDFOR

\end{algorithmic}
\end{algorithm}

\section{Experimental Setup}

\subsection{Dataset}
We use the UNSW-NB15 dataset, containing network traffic logs labeled as normal or attack behaviors across nine categories. NetFlow logs are preprocessed into token sequences representing key attributes.

\subsection{Data Preprocessing and Splitting}
The dataset is split into training (157{,}806 logs), validation (17{,}535), and test (82{,}332).

Each log entry is described by diverse features, including flow metadata (e.g., source/destination IP and ports, protocol), traffic statistics (e.g., byte and packet counts, rate, duration, throughput), temporal measures (e.g., inter-arrival time, jitter, TTL, packet loss), TCP-level indicators (e.g., round-trip time, SYN/ACK delay), and application-level/service flags. All continuous features are normalized to $[0,1]$.

Each sample has a binary label (attack or normal) and, if anomalous, one of nine attack categories: \textbf{Reconnaissance}, \textbf{Backdoor}, \textbf{DoS}, \textbf{Exploits}, \textbf{Analysis}, \textbf{Fuzzers}, \textbf{Shellcode}, \textbf{Worms}, or \textbf{Generic}. These labels support both binary and multi-class classification.

\subsection{Model Configuration}

Our system uses a dual-memory prompting architecture based on Phi-4-mini. At each inference step, the model receives a sliding window of prior log entries along with recent and persistent summaries. It outputs a natural language reasoning chain, a confidence estimate, and a classification decision.

Two memory buffers are maintained: Short-Term Memory (STM), a queue storing recent outputs; and Long-Term Memory (LTM), a persistent bank of high-confidence summaries indexed with sentence embeddings and retrieved via FAISS.

When STM reaches capacity, its contents are summarized and compressed using a fixed-size \texttt{deque}. High-confidence entries are promoted to LTM, while the rest are merged into one summary. The confidence score of the compressed summary is computed via Bayesian fusion~\cite{yu2016bayesian}. Both STM and LTM are incorporated into subsequent prompts to ensure temporal continuity and semantic grounding.

\subsection{Prompt Design}

To guide the language model in structured log analysis, we design a composite prompt with four parts: the current log, neighboring summaries, relevant long-term memory, and a task instruction block. Each part is presented sequentially, forming a structured input.

\begin{figure}[htbp]
    \centering
    \includegraphics[width=1.0\linewidth]{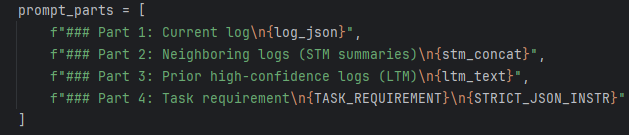}
    \caption{Prompt template sent to the language model, composed of four parts: current log, STM summaries, LTM retrievals, and task requirements.}
    \label{fig:prompt_structure}
\end{figure}

Part 1 presents the current log in raw structured form. The JSON-formatted input record contains all observed features of a single network flow.

Part 2 includes summaries of neighboring logs retrieved from the short-term memory (STM) buffer. These capture recent context that may indicate correlated or evolving anomalous behavior. Each STM entry has a natural language summary and a confidence score from earlier model output.

Part 3 adds high-confidence summaries retrieved from long-term memory (LTM) using FAISS-based similarity search. These entries represent verified anomalous cases resembling the current log, allowing the model to generalize from past examples.

Part 4 defines the task objective and enforces strict output formatting.

\begin{figure}[htbp]
    \centering
    \includegraphics[width=1.0\linewidth]{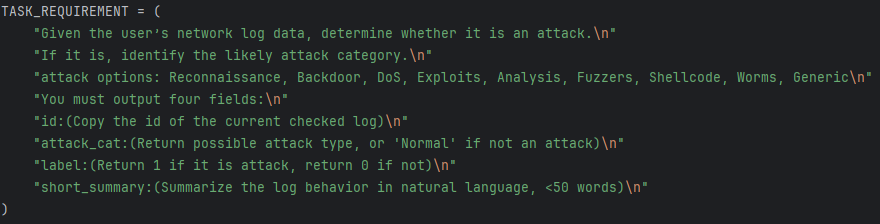}
    \caption{Instruction block provided to the language model to define the anomaly detection and classification task.}
    \label{fig:Prompt_Task}
\end{figure}

The instruction asks the model to determine whether the current log indicates an attack. If so, the model must identify the attack category from a predefined set: Reconnaissance, Backdoor, DoS, Exploits, Analysis, Fuzzers, Shellcode, Worms, and Generic. If no attack is detected, it should return \texttt{Normal}. The label is binary, where 1 indicates attack and 0 normal. Additionally, the model must provide a brief description of log behavior in fewer than 50 words.

The final output must be a valid JSON object with exactly four keys: \texttt{id}, \texttt{attack\_cat}, \texttt{label}, and \texttt{short\_summary}. No additional fields or commentary are permitted.

\begin{figure}[htbp]
    \centering
    \includegraphics[width=1.0\linewidth]{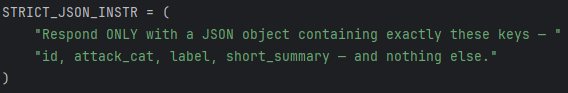}
    \caption{Instruction enforcing strict JSON output format from the LLM.}
    \label{fig:Prompt_Strict}
\end{figure}

This constraint is enforced by a final instruction string, ensuring output consistency across generations. The prompt design provides rich multi-scale context while enforcing structured responses, enabling accurate and interpretable anomaly detection.

\subsection{Comparison Methods}

We compare four methods for anomaly detection with language models:

In the \textit{zero-shot} setting, the Phi-4-mini-instruct model is directly applied without task-specific training. It receives the log and a static prompt to assign labels, evaluating its pre-trained reasoning ability.

With \textit{LoRA fine-tuning}, Phi-4-mini is adapted on labeled logs. LoRA updates a small subset of parameters by inserting trainable low-rank matrices into attention layers, reducing training cost and memory while preserving general capabilities. The fine-tuned model is then tested on unseen logs without external retrieval or prompt engineering.

For \textit{RAG}, we adopt RAG-Sequence setup, retrieving a sequence of top-K passages per step to condition generation [25] with MITRE ATT\&CK v17.1 (enterprise-attack). Following Li \textit{et al.}~\cite{li2025enhancingrag}, we emphasize semantic relevance over knowledge base size. MiniLM-L6-v2 encodes both logs and ATT\&CK definitions into embeddings, which are searched with FAISS to retrieve relevant entries. The retrieved snippets are injected into prompts and passed to Phi-4 for structured output.

Finally, our \textit{dual-memory prompting} integrates short- and long-term memories into prompts, enabling reasoning over both recent and recurring behaviors without relying on external corpora.

\subsection{Evaluation Metrics}

To evaluate our framework, we adopt standard classification metrics used in intrusion detection: \textbf{Precision}, \textbf{Recall}, \textbf{F1 Score}, and \textbf{Accuracy}. These metrics provide a comprehensive view of performance, especially under class imbalance and multi-stage attacks.

We compute the metrics using \texttt{scikit-learn}, after aligning predictions and ground truth by a unique identifier (\texttt{id}). This alignment ensures proper label correspondence.

\paragraph{Accuracy} reflects the overall proportion of correct predictions but can be misleading in imbalanced datasets.

\paragraph{Precision} measures the proportion of correctly predicted attacks among all predicted attacks, showing the system’s ability to reduce false positives.

\paragraph{Recall} measures how many actual attacks are detected among all real attack instances, critical for stealthy or staged threats.

\paragraph{F1 Score} combines precision and recall using their harmonic mean, balancing false positives and false negatives.

\subsection{Test Results}

We evaluate DM-RAG and several baselines on the UNSW-NB15 test set. Table~\ref{tab:performance} reports accuracy, precision, recall, and F1 score.

Our method achieves the highest recall (98.70\%) among models, showing strong ability to detect true positives. This indicates that DM-RAG, with its short-term memory and Bayesian fusion strategy, is more sensitive to actual attacks.

In precision (53.74\%), DM-RAG outperforms LoRA fine-tuned (44.92\%). 
However, the zero-shot Phi-4-mini has the highest precision (98.91\%), which is likely because the model is extremely conservative: it only predicts positive in very rare cases where it is highly confident. This yields very few false positives, but at the cost of an extremely low recall (0.20\%).

DM-RAG’s overall accuracy (53.64\%) is slightly lower than the MITRE-style RAG baseline (57.24\%), but its F1 score (69.59\%) is higher than all other methods, indicating balanced detection.

\begin{table}[htbp]
\caption{PERFORMANCE COMPARISON ON UNSW-NB15 TEST SET}
\centering
\begin{tabular}{lcccc}
\hline\hline
Model & Accuracy & Precision & Recall & F1 Score \\
\hline
Phi-4-mini (Zero-shot) & 45.05\% & 98.91\% & 0.20\% & 0.40\% \\
LoRA Fine-tuned & 37.97\% & 44.92\% & 46.89\% & 45.89\% \\
Phi-4 + RAG (MITRE) & 57.24\% & 68.38\% & 41.57\% & 51.70\% \\
DM-RAG (Ours) & 53.64\% & 53.74\% & 98.70\% & 69.59\% \\
\hline\hline
\end{tabular}
\label{tab:performance}
\end{table}

In summary, DM-RAG improves detection coverage (recall) while maintaining reasonable precision, suitable when missing attacks is costlier than false alarms.

\section{Limitations and Future Work}
While DM-RAG enhances reasoning and detection across long-range log sequences, several limitations remain. 

First, the heuristic promotion rule (confidence $\geq$ 0.9) may be sub-optimal, as mis-promoted summaries add noise to long-term memory (LTM), degrading inference.

Second, the method does not use explicit temporal constraints or structured log schemas, reducing precision on domain-specific data. 

Third, semantic-embedding retrieval assumes latent similarity reflects causal relevance, which may fail in noisy, imbalanced, or adversarial contexts. 

Fourth, the framework imposes no hard cap on LTM size. Each promoted summary is appended, so the FAISS index can grow indefinitely, increasing retrieval latency and GPU use. During analysis this roughly doubles runtime due to index rebuilds and queries.

Scalability is also constrained by token limits, even with the compact Phi-4-mini backbone. This may hurt performance under high-frequency events or deep behavior hierarchies. The model assumes a stationary distribution and is untested under distribution shift or continual deployment.

Future work includes schema-aware reasoning and log-template clustering for better interpretability, continual learning to prevent forgetting, and hybrid designs that fuse MITRE-style RAG with DM-RAG’s memory to improve multi-stage attack detection.

\section{Conclusion}
We presented DM-RAG, a dual-memory framework for structured log anomaly detection, enhancing temporal reasoning of compact instruction-tuned LLMs. Inspired by cognitive memory systems, DM-RAG integrates a short-term memory buffer for recent context with a FAISS-indexed long-term memory to retrieve historical patterns. 

Through memory summarization, Bayesian confidence fusion, and structured prompting, DM-RAG achieves high-recall detection of multi-stage threats while remaining lightweight and deployable.

On UNSW-NB15, DM-RAG attains the highest recall (98.70\%), with F1 of 69.59\% and precision of 53.74\%, outperforming LoRA-tuned and MITRE-style RAG baselines. The design requires no external corpora or large-scale LLMs, making it suitable for real-time, resource-constrained environments.

Beyond security, the dual-memory design generalizes to long-horizon reasoning in structured temporal data such as medical audit trails, industrial telemetry, and financial transactions. Future directions include adaptive memory management, schema-aware prompting, and continual learning for robust generalization under evolving conditions.

% \section*{Appendix: Code Availability}
% The implementation and experiment scripts are available at: \url{https://github.com/AnbiGuo/DM_RAG}

\end{document}